\documentclass[aps,12pts,showpacs,preprint]{revtex4}
\usepackage[latin9]{inputenc}
\usepackage{amsmath}
\usepackage{amssymb}
\usepackage{cancel}
\usepackage{slashed} 
\usepackage{graphicx} 
\usepackage{graphics} 
\def\b#1{\boldsymbol{#1}} 

\makeatletter
\@ifundefined{textcolor}{}
{%
 \definecolor{BLACK}{gray}{0}
 \definecolor{WHITE}{gray}{1}
 \definecolor{RED}{rgb}{1,0,0}
 \definecolor{GREEN}{rgb}{0,1,0}
 \definecolor{BLUE}{rgb}{0,0,1}
 \definecolor{CYAN}{cmyk}{1,0,0,0}
 \definecolor{MAGENTA}{cmyk}{0,1,0,0}
 \definecolor{YELLOW}{cmyk}{0,0,1,0}
}

\usepackage{bm}
\usepackage{subfigure}\usepackage{epstopdf}
\usepackage{color} 

\makeatother

\begin{document}

\title{Prediction of novel interface-driven spintronic effects}

\author{Satadeep Bhattacharjee$^{1,2~*}$, Surendra Singh$^{1}$, Dawei Wang$^{3}$, Michel Viret$^{4}$
and Laurent Bellaiche$^{1,2}$}

\affiliation{$^{1}$Physics Department,University of Arkansas, Fayetteville, Arkansas
72701, USA\\
 $^{2}$Institute for Nanoscience and Engineering, University of Arkansas,
Fayetteville, Arkansas 72701, USA\\
 $^{3}$Electronic Materials Research Laboratory, Key Laboratory of
the Ministry of Education and International Center for Dielectric
Research, Xi'an Jiaotong University, Xi'an 710049, China\\
 $^{4}$Service de Physique de l'Etat Condensé, CEA Saclay, DSM/IRAMIS/SPEC,
URA CNRS 2464, 91191 Gif-Sur-Yvette Cedex, France}
\email{sbhattac@uark.edu}
\begin{abstract}
The recently-proposed coupling between the angular momentum density
and magnetic moment {[}A. Raeliarijaona \textit{et al}, {Phys. Rev. Lett.} \textbf{110}, 137205 (2013){]} is shown here to result in the prediction of  (i) novel spin currents generated by an electrical current and (ii)   new electrical currents induced by a spin current in systems possessing specific interfaces between two different materials. Some of these spin (electrical) currents can be reversed near the interface  by reversing the applied electrical (spin) current. Similarities and differences between these novel spintronic effects and the well-known spin Hall and inverse spin Hall effects are also discussed. \end{abstract}


\maketitle
Spin transport electronics, commonly termed spintronics, is
attracting a lot of attention for fundamental purposes as well as
for its potential applications in electronic technologies \cite{Prinz,Wolf,Zutic}.
Particularly striking examples of spintronics are the spin Hall effect
(SHE) and inverse spin Hall effects (ISHE) that were, interestingly,
both first predicted \cite{D=0000D5yakonov,Hirsch} before being observed
\cite{Bakun,Tkachuk,Kato,Wunderlich}. In case of the SHE, an electric current generates a transverse spin current involving specific component of the magnetic moments, with the spin current reversing its direction when the electric current is reversed. On the other hand, it is a transverse electric current that is created and controllable by a spin-current for the ISHE. 

One may wonder if there are novel spintronic effects that remain to be discovered. For instance, it is legitimate to explore if new {\it interface-driven} spintronic phenomena may occur, once realizing that a variety of spectacular and unusual features have recently been discovered near or at the interface between two different materials.  Examples of such features, among many others, are interface-mediated conduction \cite {nature.427.423}, superconductivity \cite{science.317.1196}, multiferroic effect \cite{nmat10.753}, and improper ferroelectricity \cite {nature.452.732}. 

In particular, it is worth pursuing if the recently proposed coupling \cite{Aldo} between the angular momentum density \cite{Jackson} and magnetic moments can guide the (hypothetical) discovery of  new spintronic phenomena, since this coupling not only allowed to re-derive in  a rather straightforward fashion  the anomalous Hall effect (which is another Hall effect for which the transverse conductivity depends on the system's magnetization \cite{AHE,Nagaosa,Karplus}) but also resulted in the prediction of a novel  Hall effect \cite{AHEpaper}. Note that, in addition to the anomalous Hall effect, this coupling also explained \cite{Aldo} why magnetic vortices can be controlled by the cross product between the electric field and the magnetic field, led to the so-called
spin-current model \cite{Katsura,Dovran} in multiferroics, and yielded  the prediction of a novel anisotropic  effect \cite{Aldo}, which further emphasizes its usefulness in tackling a variety of complex problems involving electromagnetism in materials.

The goal of this manuscript is to demonstrate that, indeed, novel spintronic effects emerge near the interface between two materials from this recently proposed coupling \cite{Aldo}.

For that, let us first recall that the coupling between the angular
momentum density and magnetic moments leads to the following energy
for a conduction electron \cite{Aldo,AHEpaper}: 
\begin{equation}
{\cal E}=-\frac{a}{2}{\b r}\times({\b E}\times{\b H})\cdot{\b m}\,,\label{eqnb}
\end{equation}
where $a$ is a material-dependent constant, ${\b r}$ is the position
vector of the electron and ${\b m}$ is its magnetic moment. ${\b E}$
and ${\b H}$ are, respectively, the electric and magnetic fields
experienced by the electron.  Note that,  similar to the well-known potential energy associated with a uniform electric field, ${\cal E}$ is position-dependent while difference in energy and resulting force  (i.e., minus the gradient of ${\cal E}$ with respect to ${\b r}$) are independent of position for uniform fields.


Let us now use Eq. (\ref{eqnb}) to tackle three different cases.

\textit{Case 1}: The subscript ``C1'' will be adopted in the following to denote the energy and fields appearing in Eq. (\ref{eqnb}) for Case 1, which corresponds to the application of an electric field along a specific direction (to be chosen as the $x$-axis) to  a system having an interface (between two materials) lying in an $x-y$ plane. In that situation (denoting the unit vectors along the $x$-, $y$- and $z$-axes by ${\hat{{\bf {x}}}}$, ${\hat{{\bf {y}}}}$ and ${\hat{{\bf {z}}}}$), the electric field   ${\b E}_{C1}$ is the sum of the external electric field $E_{a}{\hat{{\bf {x}}}}$ and  a local electric field $E_{lz}{\hat{{\bf {z}}}}$ that is oriented along the $z$-axis. This local electric field occurs at the interface (see Fig. 1) and originates from the gradient of the potential across the interface.
We thus have  
\begin{equation}
{\b E}_{C1}=E_{a}{\hat{{\bf {x}}}}+E_{lz}{\hat{{\bf {z}}}}\,.\label{eqnc}
\end{equation}
Regarding the magnetic field, ${\b H}_{C1}$, we note that electrons having a magnetic moment ${\b m}$ and moving with velocity ${\b v}$ possess an electric dipole moment given by \cite{Vekstein,Panofsky}: 
\begin{equation}
{\b d}=\frac{1}{c^2} {\b v}\times {\b m}\,.\label{eqnc2}
\end{equation}
where $c$ is the speed of light. This dipole interacts with the electric field given in Eq. (\ref{eqnc}) to produce an energy of the form
\begin{equation}
{\cal E}_{int}= - {\b d} \cdot {\b E}_{C1} = - \frac{1}{c^2} ( {\b v}\times {\b m}) \cdot   {\b E}_{C1} =
- \frac{1}{c^2}  ( {\b E}_{C1} \times  {\b v} )  \cdot    {\b m} 
 \,.\label{eqnc3}
\end{equation}
This latter form can be thought as the Zeeman energy resulting from the interaction between the magnetic moment and a magnetic field given by:
\begin{equation}
{\b H}_{C1}=\frac{1}{\mu c^2} ( {\b E}_{C1} \times  {\b v} )\,, \label{eqnd}
\end{equation}
where $\mu$ is the permeability of the medium. 

Moreover, the velocity of the electron ${\b v}$
is directly proportional in magnitude, but opposite in sign, to the
applied electric field. This is because this velocity is directly
proportional but opposite in direction to the current density, the latter being simply related to the applied field via the conductivity
of the material. As a result, one can write  
\begin{equation}
{\b v}=-\frac{\sigma}{ne}E_{a}{\hat{{\bf {x}}}}\,,\label{eqnvelociy}
\end{equation}
where $\sigma$ is the conductivity, $n$ is the density of electrons
and $e$ is the magnitude of the electronic charge. This proportionality
between the velocity and the applied electric field combined with Eq. (\ref{eqnc}), allows us to rewrite Eq. (\ref{eqnd}) as  
\begin{equation}
{\b H}_{C1}=-\frac{1}{\mu c^2}\frac{\sigma}{ne}E_{a} E_{lz}{\hat{{\bf y}}}\,.\label{eqnd2}
\end{equation}
Using  Eqs. (\ref{eqnc}) and ({\ref{eqnd2}) in Eq.
(\ref{eqnb}) we obtain 
\begin{eqnarray}
{\cal E}_{C1}= & \frac{a \sigma E_{a}E_{lz}}{2ne \mu c^2}\{-x E_{a} m_{y}
+y[E_{lz}m_{z}+E_{a}m_{x}]
-zE_{lz}m_{y}\} \label{eqnb4}
 \end{eqnarray}
where ($x$, $y$, $z$) and ($m_{x}$, $m_{y}$, $m_{z}$) represent
the Cartesian components, respectively, of the position vector and
magnetic moment of the electron. It is important to realize that, unlike local electric fields associated with impurities,  $E_{lz}$ can be considered as a constant here across the (001) interface (the actual spatial variation of the field in the transition layers may depend on the quality of the interface, impurities, and other details of the transition region. To avoid getting into these unnecessary details we take the field at the interface to be constant, as consistent with the discontinuity in the potential across the interface).  Electrons near this interface will thus experience a force whose $z$-component $F_{C1,z}$  is minus the derivative of ${\cal E}_{C1}$ with respect to $z$:
\begin{equation}
F_{C1,z}=\frac{a \sigma}{2ne \mu c^2} E_{a}E_{lz}^2 m_{y} \label{eqneb}
\end{equation}

Consider now what happens when this force is incorporated into the Drude model \cite{Ashcroft} near the interface:
 \begin{equation}
\label{eqnDrude}
\frac{d  p_z }{dt}  =   -e  E_{lz} +\frac{a \sigma}{2ne \mu c^2} E_{a}E_{lz}^2 m_{y} -\frac{p_z }{\tau}\,, 
 \end{equation}
where $p_z$ is the z-component of the momentum of the electron, $e$ is the magnitude of the electronic charge, $m$ is the mass of the electron, and $\tau$ is the mean time between two successive electronic collisions. 

Multiplying this equation by $-\frac{e \tau n_I}{m}$, where $n_I$ is the density of electrons near the interface  leads to:
 \begin{equation}
\label{eqnDrude2}
j_{z,I} + \tau \frac{d  j_{z,I} }{dt}  =   \frac{n_I e^2 \tau}{m}E_{lz} - \frac{a n_I \sigma \tau}{2n m \mu c^2} E_{a}E_{lz}^2 m_{y} \,, 
 \end{equation}
 where $j_{z,I}=-\frac{e n_I}{m} p_z$ is the z-component  of the current density near the interface.
 
 The right-hand side of Eq. \eqref{eqnDrude2} (which has the dimension of a current density) has therefore two parts. The first part is given by
$\frac{n_I e^2 \tau}{m}E_{lz}$, ``simply'' arises from the existence of the local electric field inherent to the interface, and is experienced by any electron independently of the sign of its $m_{y}$. On the other hand, the second part has a {\it global sign} that depends on the sign of $m_{y}$:
\textit{in the case that the coefficient $a$  is positive}, electrons having a positive $m_{y}$ will 
generate a negative z-component of the current density near the interface while electrons possessing a negative $y$-component of their magnetic moment will result in a positive (opposite) contribution to $j_{z,I}$ when the applied field $E_{a}$ is along the $+x$-axis (the opposite situation holds if $a$ is \textit{negative}). As a result, the (001) interface tends to act as a ``filter'' to deflect electrons with a  definite sign for the $y$-component of their magnetic moment along a specific side of the interface and a spin current exists along the z-axis, provided that the magnitude of the second term of Equation \eqref{eqnDrude2} is not negligible in front of the first term \cite{foonotesteadystate}. Interestingly,  motions that are precisely opposite to the ones we just described (related to the second term of  Eq. \eqref{eqnDrude2}) 
will occur when the applied field is reversed from the $+x$ to $-x$ direction, because Eq. \eqref{eqneb} is linearly dependent on $E_{a}$. 
Such field-controllable motion of electrons with specific magnetic moment, shown schematically in Figs. 1, resembles the  spin Hall effect (SHE) \cite{D=0000D5yakonov,Hirsch,Bakun,Tkachuk,Kato,Wunderlich,Dyakonov2,Sinova,Guo1,Guo2}.
Note, however, that the SHE is different from the effect described here because (1) SHE can occur in bulk systems while the  effect predicted here is an interface-driven phenomenon; and (2), unlike the new effect, SHE does not require the coupling between angular momentum density and magnetic moments for occurring. For instance, in case of the extrinsic SHE \cite{Dyakonov2}, which is typically explained in terms of  the so-called  skew-scattering and side-jump mechanisms involving impurities \cite{Mott1,Mott2,Goldberger,Karplus,Smit,Luttinger,Berger,Lyo}, (i) one can just consider the Zeeman interaction energy between the magnetic moments and the magnetic field given by  Eq. \eqref{eqnd}; (ii) realize that the local field around an impurity is {\it inhomogeneous}, unlike the case considered here; and (iii) finally obtain a force along the $z$-axis by taking the derivative of this Zeeman energy with respect to $z$,  viz.,
$-\frac{\sigma}{2ne \mu c^2} E_{a} \frac{\partial E_{lz}}{\partial z} m_{y}$. This force naturally explains the extrinsic SHE in bulk systems, since one can easily demonstrate that $\frac{\partial E_{lz}}{\partial z}$ has a definite sign around an impurity (e.g., by considering a central potential whose gradient provides the local electric field).

\textit{Case 2}:  Let us now consider Case 2, which is similar to Case 1 with the important exception that the interface now lies in a (011) plane. 
Using the subscript ``C2'' to denote appropriate quantities, the electric and magnetic fields of Eq. (\ref{eqnc}) and (\ref{eqnd}) in this case become
\begin{equation}
{\b E}_{C2}=E_{a}{\hat{{\bf {x}}}}+E_{ly}{\hat{{\bf {y}}}}+E_{lz}{\hat{{\bf {z}}}}\,\label{eqnc2ndcase}
\end{equation}
and
\begin{equation}
{\b H}_{C2}=\frac{1}{\mu c^2}\frac{\sigma}{ne}E_{a}(E_{ly}{\hat{{\bf z}}}-E_{lz}{\hat{{\bf y}}})\,.\label{eqnd2ndcase}
\end{equation}
Note that the local electric field in this case acquires a $y$-component, $E_{ly}$ (equal in magnitude to $E_{lz}$) since we are now dealing with a (011) interface.
Then the energy associated with the coupling between the angular momentum density and magnetic moments can be written as:
\begin{align}
{\cal E}_{C2}= & \frac{a \sigma E_{a}}{2ne \mu c^2}\{x E_{a}(E_{ly}m_{z}-E_{lz}m_{y})\}\label{eqnb42ndcase}\\
 & +\frac{a \sigma E_{a}}{2ne\mu  c^2}\{y[\{(E_{ly})^{2}+(E_{lz})^{2}\}m_{z}+E_{a}E_{lz}m_{x}]\}\nonumber \\
 & -\frac{a\sigma E_{a}}{2ne \mu c^2}\{z[\{(E_{ly})^{2}+(E_{lz})^{2}\}m_{y}+E_{a}E_{ly}m_{x}]\}\,.\nonumber 
\end{align}
The $y$- and $z$-components of the force, $F_{C2,y}$ and $F_{C2,z}$,
associated with this energy near the interface are then:
\begin{subequations}
\begin{align}
 & F_{C2,y}=-\frac{a \sigma}{2 ne\mu c^2}E_{a}\left(E_{ly}^{2}+E_{lz}^{2}\right)m_{z}-\frac{a\sigma}{2 ne\mu c^2}E_{a}^{2}E_{lz}m_{x}\,,\label{eqnea2ndcase}\\
 & F_{C2,z}=\frac{a \sigma}{2 ne\mu c^2}E_{a}\left(E_{ly}^{2}+E_{lz}^{2}\right)m_{y}+\frac{a\sigma}{2 ne\mu c^2}E_{a}^{2}E_{ly}m_{x}\,.\label{eqneb2ndcase}
\end{align}
\end{subequations}

Incorporating these two components into the Drude model \cite{Ashcroft} near the interface yields:
 \begin{subequations}
 \begin{align}
& \frac{d  p_y }{dt}  =    -e  E_{ly} -\frac{a \sigma}{2 ne\mu c^2}E_{a}\left(E_{ly}^{2}+E_{lz}^{2}\right)m_{z}-\frac{a\sigma}{2 ne\mu c^2}E_{a}^{2}E_{lz}m_{x} -\frac{p_y}{\tau} \label{eqnDrudeCase2a}\\
& \frac{d  p_z }{dt}   =    -e  E_{lz} +\frac{a \sigma}{2 ne\mu c^2}E_{a}\left(E_{ly}^{2}+E_{lz}^{2}\right)m_{y}+\frac{a\sigma}{2 ne\mu c^2}E_{a}^{2}E_{ly}m_{x} -\frac{p_z}{\tau}\,.\label{eqnDrudeCase2b}
\end{align}
 \end{subequations}

Multiplying these equations by $-\frac{e \tau n_I}{m}$ results in:
 \begin{subequations}
  \begin{align}
& j_{y,I}  + \tau \frac{d  j_{y,I} }{dt} =    \frac{n_I e^2 \tau}{m}E_{ly} +\frac{a n_I \sigma \tau}{2 nm\mu c^2}E_{a}\left(E_{ly}^{2}+E_{lz}^{2}\right)m_{z}+\frac{a n_I \sigma \tau}{2 n m \mu c^2}E_{a}^{2}E_{lz}m_{x}
\label{eqnDrude2Case2a}\\
& j_{z,I}  + \tau \frac{d  j_{z,I} }{dt} =    \frac{n_I e^2 \tau}{m}E_{lz}-\frac{a n_I \sigma \tau }{2 n m \mu c^2}E_{a}\left(E_{ly}^{2}+E_{lz}^{2}\right)m_{y}-\frac{a n_I \sigma \tau}{2 n m \mu c^2}E_{a}^{2}E_{ly}m_{x}  \,, 
\label{eqnDrude2Case2b}
\end{align}
 \end{subequations}
 where $j_{y,I}$ and $j_{z,I}$ are the y- and z-components of the current density near the interface.
 
The first terms on the right-hand side of Eq. \eqref{eqnDrude2Case2a} and  \eqref{eqnDrude2Case2b} indicate that the local electric field existing near the (011) interface induces a current density that is oriented along  the [011] direction, independently of the sign and magnitude of the Cartesian components of the magnetic moments  of the electrons. 
If we assume that $a$ is positive, the second terms on the  right-hand side of Eq. \eqref{eqnDrude2Case2a} and  \eqref{eqnDrude2Case2b} tell us that that the electrons with positive  (negative) $m_z$ tends to generate an additional positive (negative) y-component of the current density near the interface  while those with positive  (negative) $m_y$ induces an additional negative (positive)  z-component of the current density near the interface when the electric field is applied along the $+x$-axis, with these motions reversing when the applied electric is reversed. In other words, as in Case 1, the presence of the interface leads to  spin currents that are controllable by the applied electric field. What is new in Case 2 with respect to Case 1 is the presence of the third terms on the  right-hand side of Eq. \eqref{eqnDrude2Case2a}
and  \eqref{eqnDrude2Case2b}. These third terms indicate that,  if $a$ is positive, electrons with positive  $m_x$ will generate another current density (near the interface) that is   along $+\hat{{\bf {y}}}-\hat{{\bf {z}}}$, that is, along a direction that is {\it parallel} to the interface, while electrons with negative $m_x$ will generate an opposite additional current density. Interestingly, this ``in-plane'' spin-current involving the $x$ component of the magnetic moments can not be reversed  when the applied electric field is reversed, because the third terms of Eq. \eqref{eqnDrude2Case2a} and  \eqref{eqnDrude2Case2b}  are proportional to the \textit{square} of the applied electric field. This new (interface-driven) phenomenon, different in nature from a spin Hall-like effect,  is summarized in Fig. 2 \cite{foonotesteadystate}.

\textit{Case 3}: Let us now consider the situation (for which the  subscript ``C3'' will be used in the following) of a spin current in which electrons
having a negative $m_z$ move along the $+y$ axis with a velocity $v_{y}{\hat{{\bf {y}}}}$ while precisely the opposite motion occurs for electrons having a positive $m_{z}$.
For simplicity, we do not include here the possibility that these
spin electrons may also have a definite $x-$ or $y-$components of their magnetic moments. As shown in Fig. 3, an interface lying in a $y-z$ plane is  considered in this system, which results in the formation of  local electric field aligned along the $x$-axis at this interface:
\begin{equation}
{\b E}_{C3}=E_{lx}{\hat{{\bf {x}}}}.\label{eqncISHE}
\end{equation}
Because of the existence of the spin current,
the electrons having a \textit{negative} z-component of their magnetic
moments will experience  a magnetic
field of the form  
\begin{equation}
{\b H}_{C3,\downarrow}=\frac{1}{\mu c^2}{\b E}_{C3} \times v_{y}{\hat{{\bf {y}}}}=\frac{1}{\mu c^2}v_{y}E_{lx}{\hat{{\bf z}}}\,,\label{eqndISHEn}
\end{equation}
while electrons possessing a \textit{positive} $m_{z}$ will feel
 precisely the \textit{opposite} magnetic field
\begin{equation}
{\b H}_{C3,\uparrow}=-\frac{1}{\mu c^2}{\b E}_{C3} \times v_{y}{\hat{{\bf {y}}}}=-\frac{1}{\mu c^2}v_{y}E_{lx}{\hat{{\bf z}}}\,.\label{eqndISHEp}
\end{equation}
Inserting Eqs. (\ref{eqncISHE}) and ({\ref{eqndISHEn}) into Eq. (\ref{eqnb}) gives for electrons with \textit{negative} $m_{z}$,
\begin{equation}
{\cal E}_{C3}=-\frac{a v_{y}}{2 \mu c^2} x E_{lx}^{2}|m_{z}|\,,\label{eqnb3ISHEm}
\end{equation}
where $|m_{z}|$ is the absolute value of the $z$-component of the
magnetic moment. Interestingly, the electrons having a \textit{positive}
$m_{z}$ also possess  the same energy, ${\cal E}_{C3}$ (this
can be shown either by inserting Eqs. (\ref{eqncISHE}) and ({\ref{eqndISHEp})
into Eq. (\ref{eqnb}) or by realizing that Eq. (\ref{eqnb}) is invariant
under  a simultaneous change of sign of the magnetic field and magnetic moment).
As a result, \textit{all} electrons experience the same force along the $x$-axis near the interface:
\begin{equation}
 F_{C3,x}=\frac{av_{y}}{2\mu  c^2} E_{lx}^{2}|m_{z}|\,.\label{eqneISHEa}
\end{equation}
Consider this force in the Drude model \cite{Ashcroft} near the interface therefore gives:
 \begin{equation}
\label{eqnDrudeCase3}
\frac{d  p_x }{dt}  =   -e  E_{lx} +\frac{av_{y}}{2\mu  c^2} E_{lx}^{2}|m_{z}| -\frac{p_x }{\tau}\,, 
 \end{equation}

Multiplying this latter equation by $-\frac{e \tau n_I}{m}$ leads to:
 \begin{equation}
\label{eqnDrude2Case3}
j_{x,I} + \tau \frac{d  j_{x,I} }{dt} =   \frac{n_I e^2 \tau}{m}E_{lx} -\frac{a n_I e v_{y} \tau}{2 m \mu  c^2} E_{lx}^{2}|m_{z}|  \,, 
 \end{equation}
 where $j_{x,I}=-\frac{e n_I}{m} p_x$ is the x-component  of the current density near the interface.
Equation (\ref{eqnDrude2Case3}) therefore tells us that, in addition to $\frac{n_I e^2 \tau}{m}E_{lx}$ that is valid for any electron,  electrons with both positive and negative $m_{z}$ will generate an additional negative x-component of current density in the vicinity of the interface if $a$ is positive (otherwise this additional x-component is positive) thereby creating an additional \textit{electrical}
current that is transverse to the spin current. This additional transverse electrical
current, unlike $\frac{n_I e^2 \tau}{m}E_{lx}$, can be reversed by switching the direction of the spin current  since Eq. (\ref{eqneISHEa}) depends linearly  on the spin-current
velocity $v_{y}$. All these effects, shown schematically in Figs. 3, are reminiscent of the so-called inverse spin Hall effect (ISHE) \cite{D=0000D5yakonov,Hirsch}. However, unlike the traditional ISHE, the effect predicted here is interface-driven and originates from the coupling
between angular momentum density and magnetic moments. 

Let us also indicate what happens in the {\it steady state} regime. In that situation, $ \frac{d  j_{x,I} }{dt}$=0 and $j_{x,I}$ will vanish as a result of the formation of another electric field along the x-axis, arising from the transfer of charge associated with the electrons that  have already crossed the interface and that will oppose the further motion of electrons along the x-axis (as, e.g., similar to the cases of the regular Hall effect or p-n junction \cite{Ashcroft}). When denoting such new electric field as $E_{opp,x}$, Eq. (\ref{eqnDrude2Case3}) will then lead to
 \begin{equation}
\label{eqnDrude2Case3steadystate}
 E_{opp,x} =   -E_{lx} +\frac{a  v_{y} }{2 e \mu  c^2} E_{lx}^{2}|m_{z}|  \,, 
 \end{equation}

Interestingly, Eq. (\ref{eqnDrude2Case3steadystate}) therefore tells us that $E_{opp,x}$ (and the resulting associated voltage existing inside the system along the x-axis)  will change in magnitude when reverting $v_{y}$, which can be practically used to experimentally demonstrate the existence of the (novel) second term of Eq. (\ref{eqnDrude2Case3}).

Note that this interface-driven formation of a voltage from the application of a spin current is similar in nature from the recently observed spin-to-charge conversion at the interface between non-magnetic materials  \cite{Rojas}. However,  these two effects are technically different because the latter phenomenon (which is termed the inverse Rashba-Edelstein effect \cite{Edelstein}) involves the application of a spin current that is {\it along} (rather than perpendicular to) the normal of the interface and the creation of an electrical current that is perpendicular (rather than parallel) to the normal of the interface. In fact, it is interesting to realize that the experiment of Ref.  \cite{Rojas} precisely corresponds to the {\it reciprocal} situation 
of our `Case 1':  In Ref. \cite{Rojas}, a spin current
flowing along the normal of the interface (i.e., the z-axis) and involving spins oriented along an axis that lies inside the interface plane (e.g., spins having y-components) leads to the creation of an electric voltage that is along the third Cartesian axis (e.g., the x-axis), while our Case 1 is about the formation of a spin current oriented along the z-axis (which is parallel to the normal of the interface) of spins having y-components from the application of an electric field being along the x-axis.

Finally, one can easily demonstrate, that the case that differs from``Case 3'' only by having an interface along a (110) (rather than a (100) plane), will also result in the creation of an additional electric current along the normal of the interface (in the non-steady state situation), which can be  reversed by reversing the spin current.

In summary, we have demonstrated that the coupling between angular
momentum density and magnetic moments predicts novel spintronic effects near interfaces between two different materials. We hope that these novel effects, summarized in Figs. 1-3, will stimulate experimental work for their conformation.  Note that, as hinted in Ref. \cite{AHEpaper} and in the supplemental material of Ref. \cite{Aldo}, the $a$ coefficient appearing in Eq. (\ref{eqnb}) requires spin-orbit interactions to be non-zero. As a result,  the phenomena predicted here are more likely to be observed in systems with strong spin-orbit interactions. It may also be worth investigating if theories involving Berry-phase curvature \cite{Sinova,Guo1,Guo2} can also explain the novel phenomena predicted here.  This appears to be a promising  scenario especially when we recall 
that the coefficient $a$ of Eq.(\ref{eqnb}) has been recently found \cite{AHEpaper} to be directly related to Berry-phase curvature \cite{Berry} in case of the anomalous Hall effect.

We thank ARO Grant No. W911NF-12-1-0085 for personnel support. Office of Basic Energy Sciences, under contract ER-46612, ONR Grants No.
N00014-11-1-0384 and N00014-12-1-1034, and NSF Grant No. DMR-1066158
are also acknowledged for discussions with scientists sponsored by
these grants. The authors thank Vincent Cross and Huaxiang Fu for
insightful discussions.

\newpage{}

\newpage{}
\begin{figure}
\includegraphics[width=135mm,height=135mm]{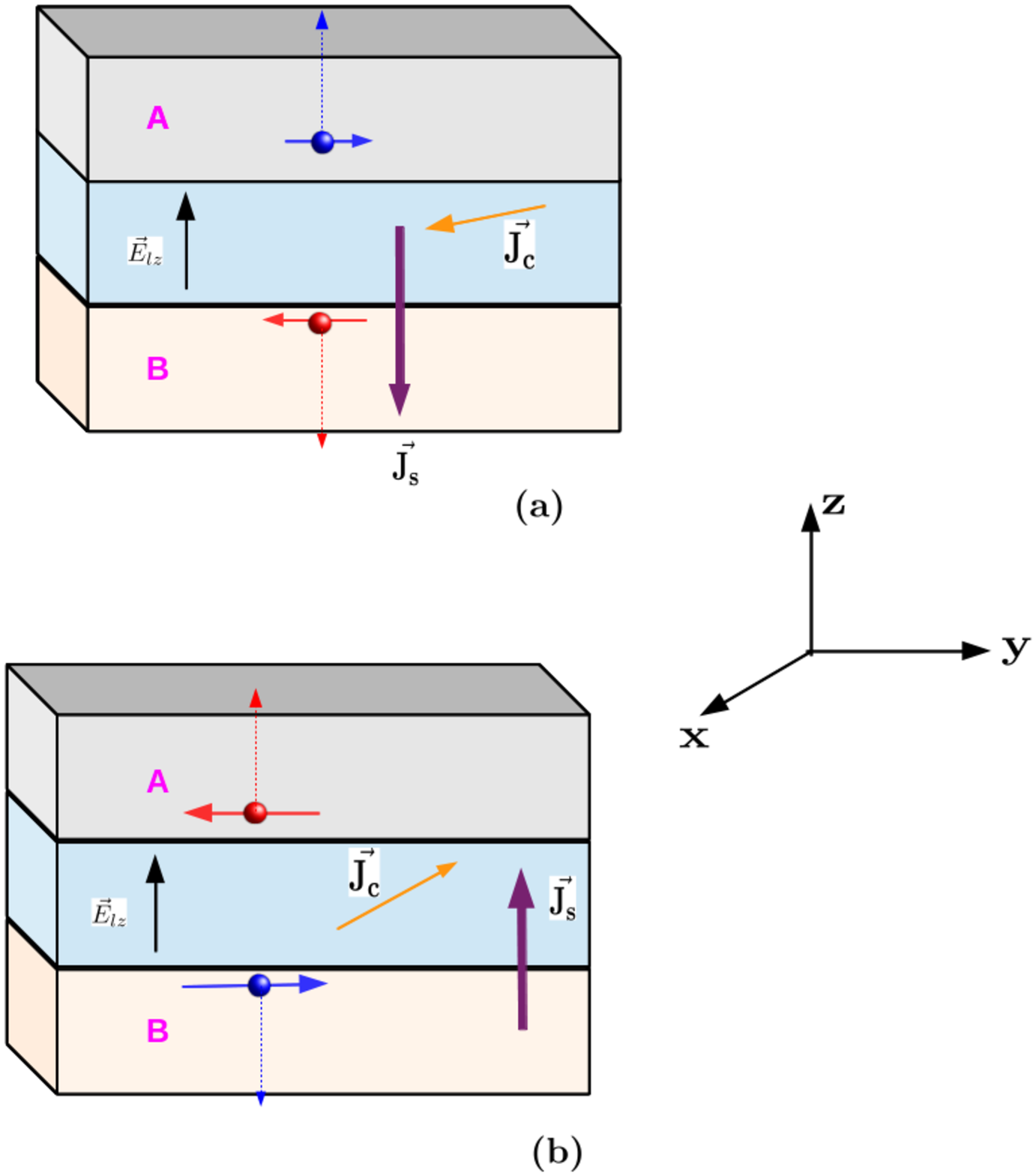}
\caption{(Color online) Schematics of the predicted effect associated with
Case 1 for a system having a (001) interface between two materials. Panel (a) and (b) differ by the direction of the applied electric field ${\b E}_{a}$ (i.e., parallel or antiparallel to the $x$-axis), and show the resulting electric-field-controllable transverse motion of the $y$-component of the magnetic moments along the $z$-axis near the interface (associated with the second term of
 Eq. \eqref{eqnDrude2}) -- resulting in a spin current, ${\b J}_{s}$. Electrons are represented by dots while the components of their magnetic moments are shown by solid lines going though the dots. The direction of the force experienced by these electrons (associated with
  Eq. \eqref{eqneb}) is displayed via dashed lines.}

\label{Fig.1} 
\end{figure}

\vspace{5mm}

\begin{figure}
\includegraphics[width=135mm,height=135mm]{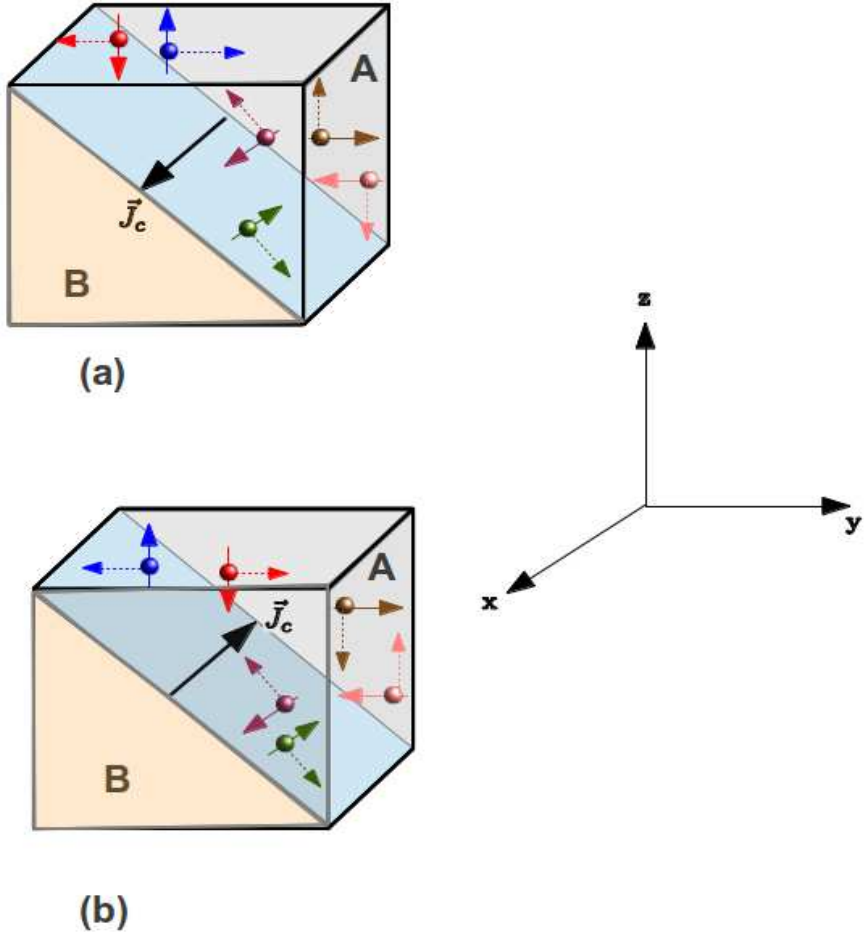}
\caption{(Color online) Schematics of the predicted effect associated with
Case 2 for a system possessing a (011) interface between two materials. Panel (a) and (b) differ by the direction of the applied electric field ${\b E}_{a}$  (i.e., parallel or antiparallel to the $x$-axis). They show the corresponding motion of all components of the magnetic moments near the interface (associated with the second and third terms of Eqs. \eqref{eqnDrude2Case2a} and  \eqref{eqnDrude2Case2b}). Electrons are represented by dots while the components of their magnetic moments are shown by solid lines going though the dots. The direction of the force (associated with Eqs. \eqref{eqnea2ndcase} and \eqref{eqneb2ndcase})
experienced by these electrons is displayed via dashed lines.}

\label{Fig.2} 
\end{figure}

\vspace{5mm}

\begin{figure}
\includegraphics[width=125mm,height=125mm]{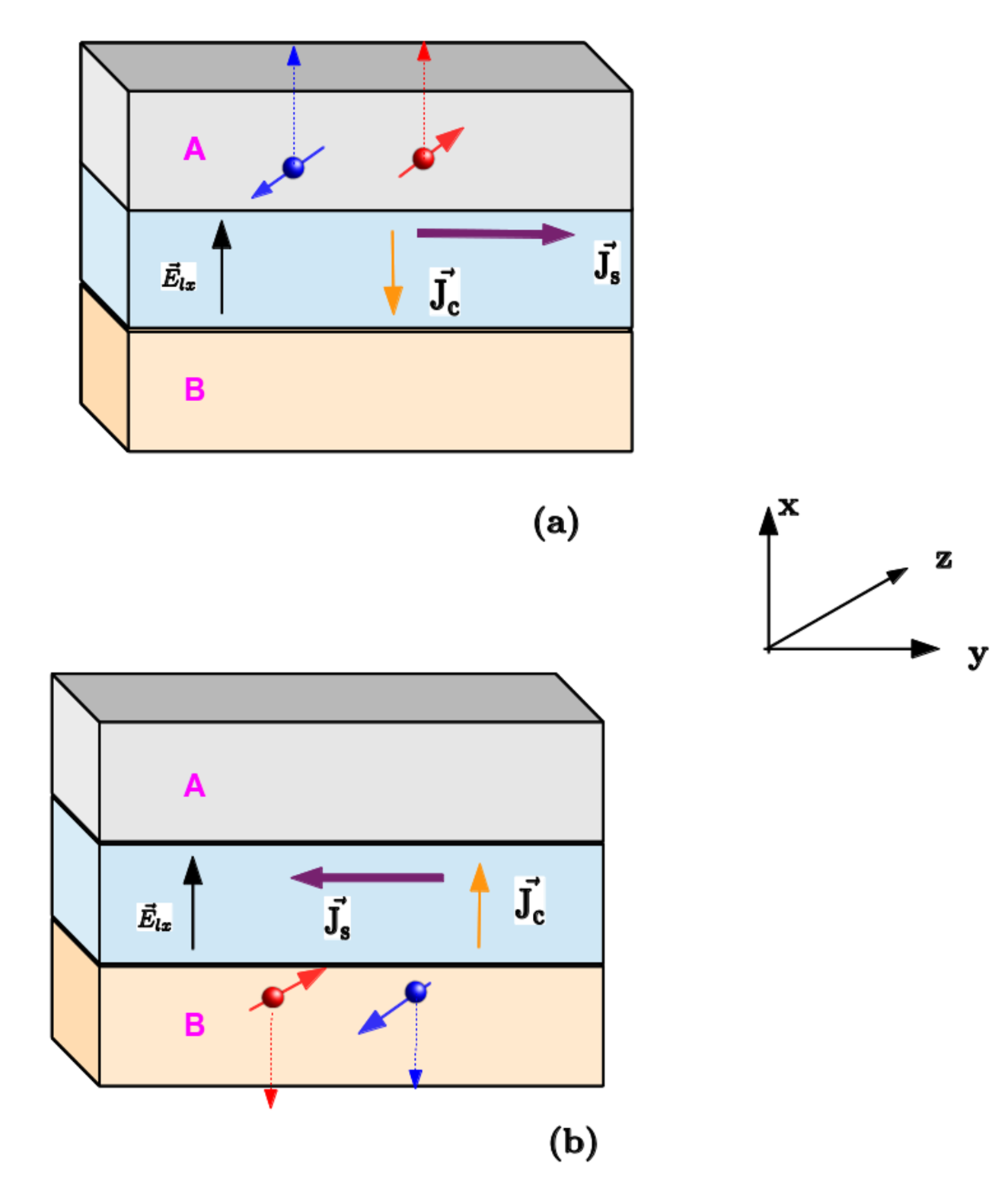}
\caption{(Color online) Schematics of the predicted effect associated with
Case 3 for a system having a (100) interface between two materials. Panel (a) and (b) differ by 
the direction of the spin current  ${\b J}_{s}$  (parallel or antiparallel  to the $y$-axis),
and  show the resulting formation of an electrical current ${\b J}_{c}$ (associated with the second term of Eq. (\ref{eqnDrude2Case3}))
along the normal of the interface. Electrons are represented by dots while the components of their magnetic moments are shown by solid lines going though the dots. The direction of the force experienced by these electrons (and associated with Eq. (\ref{eqneISHEa}))
is displayed via dashed lines}

\label{Fig.2} 
\end{figure}


}}}

\begin{thebibliography}{10}
\bibitem{foonotesteadystate} Note that Eqs.  (\ref{eqnDrude2}), (\ref{eqnDrude2Case2a}) and (\ref{eqnDrude2Case2b})  are for non-steady state situations. When the steady state will set in,  the current density solely generated by the local electric field existing near the interface (i.e, $\frac{n_I e^2 \tau}{m}E_{lz}$ in Eqs. (\ref{eqnDrude2}) and (\ref{eqnDrude2Case2b}), and $\frac{n_I e^2 \tau}{m}E_{ly}$ in Eq. (\ref{eqnDrude2Case2a})) should be counterbalanced by the electric field created by the charge imbalance induced by the electrons that have crossed such interface. Moreover, in the steady state situation, spins imbalance should be found at the two sides of the interface in Cases 1 and 2.
\end{thebibliography}
\end{document}